\documentclass{jnmp}
\usepackage{amsmath}

\JNMPnumberwithin{equation}{section}
\newtheorem{theorem}{Theorem}
\newtheorem{proposition}{Proposition}
\newtheorem{lemma}{Lemma}
\theoremstyle{definition}
\newtheorem{remark}{Remark}

\begin{document}
\renewcommand{\evenhead}{K Kajiwara}
\renewcommand{\oddhead}{$q$-Painlev\'e III Equation}
\thispagestyle{empty}

\FirstPageHead{10}{3}{2003}{\pageref{kajiwara-firstpage}--\pageref{kajiwara-lastpage}}{Article}

\copyrightnote{2003}{K Kajiwara}

\Name{On a $\boldsymbol{q}$-Difference Painlev\'e III Equation:
II. Rational Solutions}
\label{kajiwara-firstpage}

\Author{Kenji KAJIWARA}

\Address{Graduate School of Mathematics, Kyushu
University, 6-10-1 Hakozaki,\\
Higashi-ku, Fukuoka 812-8512, Japan\\
E-mail: kaji@math.kyushu-u.ac.jp}

\Date{Received July 26, 2002; Revised November 18, 2002;
Accepted November 23, 2003}

\begin{abstract}
\noindent
Rational solutions for a $q$-difference analogue of the Painlev\'e III
equation are consi\-de\-red. A Determinant formula of Jacobi--Trudi type for the
solutions is constructed.
\end{abstract}

\section{Introduction}
This paper is the second half of the work on a $q$-difference
Painlev\'e III equation ($q$-P$_{\rm III}$),
\begin{gather}
 \overline{f_1}=\frac{q^{2N+1}c^2}{f_0f_1}
\; \frac{1+a_0q^nf_0}{a_0q^n+f_0} ,\nonumber\\
 \underline{f_0}=\frac{q^{2N+1}c^2}{f_0f_1}
\; \frac{a_1q^{-n+\nu}+f_1}{1+a_1q^{-n+\nu}f_1} ,
\label{qP3}
\end{gather}
where $f_i=f_i(n;\nu,N)$ $(i=0,1)$ are dependent variables,
$n\in\mathbb{Z}$ is the independent variable, $\nu$,$N\in\mathbb{Z}$ are
parameters, and $q$, $a_0$, $a_1$, $c$ are constants. Moreover,
$\overline{f_i}$ and $\underline{f_i}$ denote $f_i(n+1;\nu,N)$ and
$f_i(n-1;\nu,N)$, respectively.

In the previous paper~\cite{KK:qp3}, we have discussed the derivations,
symmetry and particular solutions of Riccati type of $q$-P$_{\rm III}$
(\ref{qP3}).  In this paper, we consider the rational solutions.

We remark that $q$-P$_{\rm III}$ (\ref{qP3}) appears as a dynamical
system on certain rational surface (``Mul.6 surface'' in Sakai's
classification~\cite{Sakai}). Moreover, $q$-P$_{\rm III}$ admits a
symmetry of (extended) affine Weyl group of type $A_1^{(1)}\times
A_1^{(1)}$ as a group of B\"acklund transformations, which is the same
type as that for the Painlev\'e III equation (P$_{\rm III}$).

The six Painlev\'e equations (P$_{\rm J}$, ${\rm J}={\rm I},{\rm II},\ldots,{\rm VI}$), except for
P$_{\rm I}$, admit two classes of classical solutions for
special values of parameters; one is so-called transcendental classical
solutions, namely, one-parameter family of particular solutions
expressible in terms of classical special functions of hypergeometric
type. Another class is algebraic or rational solutions.

There are two classes among algebraic or rational solutions. One class
consists of such solutions that can be regarded as special case of
transcendental classical solutions.  Another class of solutions are
expressed in terms of log derivative or ratio of certain characteristic
polynomials (after some change of independent variable), and satisfy
recurrence relations of Toda type. We call such polynomials ``special
polynomials'' for the Painlev\'e equations. They are sometimes referred
as Yablonskii--Vorob'ev polynomials for P$_{\rm II}$, Okamoto polynomials
for P$_{\rm IV}$, Umemura polynomials for P$_{\rm III}$, P$_{\rm V}$ and
P$_{\rm VI}$.

It may be natural then to ask a question: what are those special
polynomials? One answer is that the special polynomials are
specialization of the Schur function or its generalizations. Such
description has been established through the Jacobi--Trudi type
determinant formulas for the special polynomials. Moreover, the entries
are expressed in terms of classical special polynomials
\cite{KM:p3,KO:p2,KO:p4,MOK:p5,M:p6,NY:p4,NY:p5ume,Y:det}.

Compared to the continuous case, the amount of known results for discrete
Painlev\'e equations is not so much.  Although it is not difficult to
construct rational solutions if we have B\"acklund transformations of given
discrete Painlev\'e equation (a systematic method to construct B\"acklund
transformations is given in~\cite{JRB}), construction of determinant
formulas and identifying the special polynomials as classical object
sometimes contain technical difficulty.  So far the rational solutions
are systematically discussed for the standard discrete Painlev\'e II
equation~\cite{KOY:dp2}, a $q$-difference Painlev\'e IV
equation~\cite{KNY:qp4,KNY:qKP} and a $q$-difference Painlev\'e V
equation~\cite{M:qp5} (determinant formula for rational solutions of
standard discrete Painlev\'e IV equation is conjectured in~\cite{JK:dp4}).

The purpose of this paper is to construct a class of rational solutions
for $q$-P$_{\rm III}$ (\ref{qP3}) and present a determinant formula of
Jacobi--Trudi type.

This paper is organized as follows. In Section 2 we give a brief review
for rational solutions of P$_{\rm III}$ and show that all the rational
solutions admit determinant formula of Jacobi--Trudi type. In Section~3
we construct a class of rational solutions for $q$-P$_{\rm III}$ and
present the Jacobi--Trudi type determinant formula, which is the main
result of this paper. We give the proof in Section~4.

\section{Rational solutions of P$\boldsymbol{{}_{\rm III}}$}

It is known that the Painlev\'e III equation (P$_{\rm III}$),
 \begin{equation}
  \frac{d^2v}{dx}=\frac{1}{v}\left(\frac{dv}{dx}\right)^2 - \frac{1}{x}\frac{dv}{dx}
+\frac{1}{x}\left(\alpha v^2+\beta\right)
+ \gamma v^3+ \frac{\delta}{v},\qquad \gamma=-\delta=4,\label{p3}
\end{equation}
admits a class of rational solutions expressible in terms of
determinant of Jacobi--Trudi type associated with the
two-core partition $\lambda=(N,N-1,\ldots,1)$.

\begin{theorem}[\cite{KM:p3}]\label{theorem:p3ume}
 Let $p_k(x,s)$, $k\in \mathbb{Z}$ be polynomials in $x$ defined by
\begin{equation}\label{Laguerre}
 \sum_{k=0}^\infty p_k(x,s)\lambda^k = (1+\lambda)^{s+1/2}{\rm e}^{2x\lambda},\quad
k\geq 0,\qquad p_k(x,s)=0,\quad k<0.
\end{equation}
For each $N\in\mathbb{Z}_{\geq 0}$, we define a polynomial $U_N(x,s)$
by
\begin{equation}\label{p3tau}
 U_N(x,s)=\left|\begin{array}{cccc}
       p_N(x,s)&p_{N+1}(x,s) &\cdots &p_{2N-1}(x,s) \\
       p_{N-2}(x,s)&p_{N-1}(x,s) &\cdots &p_{2N-3}(x,s) \\
\vdots & \vdots &\ddots &\vdots\\
       p_{-N+2}(x,s)&p_{-N+3}(x,s) &\cdots &p_{1}(x,s) \\
        \end{array}\right|,\qquad U_0=1.
\end{equation}
Then,
\begin{equation}
 v = \frac{U_{N}(x,s-1)U_{N-1}(x,s)}{U_N(x,s)U_{N-1}(x,s-1)},
\end{equation}
satisfies P$_{\rm III}$ (\ref{p3}) with $\alpha=4(s+N)$, $\beta=4(-s+N)$.
\end{theorem}

This theorem asserts that a class of rational solutions of P$_{\rm III}$
is given by ratio of specialization of the Schur functions associated
with partition $\lambda=(N,N-1,\ldots,1)$.  We note that $p_k(x,s)$ is
nothing but the Laguerre polynomial $L_k^{(s+1/2-k)}(-2x)$. The
polynomials $U_N(x,s)$ (after some rescaling) are sometimes called the
Umemura polynomials for P$_{\rm III}$.

It may not be meaningless to state here that the above solutions
essentially cover all the rational solutions of P$_{\rm III}$.
Classification of rational solutions for P$_{\rm III}$ is discussed
in~\cite{Gromak3,Murata:p3}.  According to~\cite{Murata:p3}, the result is
summarized as follows.

\begin{theorem}[\cite{Murata:p3}]\label{theorem:Murata}
1. Let $\alpha=-4\theta_\infty$ and $\beta=4(\theta_0+1)$ in P$_{\rm III}$
(\ref{p3}).  Then P$_{\rm III}$ admits rational solutions if and only if
there exists an integer $I$ such that (i) $\theta_\infty+\theta_0+1=2I$
or (ii) $\theta_\infty-\theta_0-1=2I$.

2. If P$_{\rm III}$ has rational solutions, then the number of
rational solutions for each ($\theta_\infty$,$\theta_0$) is two or
four. Furthermore, P$_{\rm III}$ admit four rational solutions if and
only if there exists two integers $I$ and $J$ such that
$\theta_\infty+\theta_0+1=2I$ and $\theta_\infty-\theta_0-1=2J$.
\end{theorem}

Obviously, the solutions in Theorem~\ref{theorem:p3ume} correspond to the
case (ii) with $I\leq 0$, since in this case $(\theta_\infty,\theta_0)$
is given by $(\theta_\infty,\theta_0)=(-s-N,-s+N-1)$ with
$N\in\mathbb{Z}_{\geq 0}$. We next remark that if we define $U_N$
$(N<0)$ by
\begin{equation}\label{negative}
 U_N = (-1)^{N(N+1)/2} U_{-N-1},
\end{equation}
then Theorem~\ref{theorem:p3ume} is extended to all $N\in \mathbb{Z}$. In
fact, Theorem~\ref{theorem:p3ume} was proved based on the fact that if $U_N$
satisfies the Toda equation,
\begin{gather} (2N+1)U_{N+1}U_{N-1}=-\frac{x}{2}
\left[\frac{d^2U_N}{dx}^2 U_N-\left(\frac{dU_N}{dx}\right)^2
\right]\nonumber\\
\label{p3Toda}
\phantom{(2N+1)U_{N+1}U_{N-1}=}{}
- \frac{1}{2}\frac{dU_N}{dx}\; U_N+\left(2x+s+\frac{1}{2}\right)U_N^2,
\end{gather}
with the initial condition,
\begin{equation}
 U_{-1}=U_0=1,
\end{equation}
then $v = \frac{U_{N}(x,s-1)U_{N-1}(x,s)}{U_N(x,s)U_{N-1}(x,s-1)}$
satisfies P$_{\rm III}$ for $N\in \mathbb{Z}_{>0}$. This fact is easily
generalized to $N\in\mathbb{Z}$ by extending $U_N$ according to the Toda
equation (\ref{p3Toda}) to $N\in \mathbb{Z}_{<0}$.  Therefore, the rational
solutions in Theorem~\ref{theorem:p3ume} with the
extension~(\ref{negative}) cover one class of rational solutions of the case~(ii).

According to Theorem~\ref{theorem:Murata}, there must be another class of
rational solutions for the case~(ii). However, noticing that P$_{\rm III}$
(\ref{p3}) admits a B\"acklund transformation,
\begin{equation}
 (\theta_\infty,\theta_0)\mapsto (\overline{\theta}_\infty,\overline{\theta}_0)
=(-\theta_0-1,-\theta_\infty-1),\qquad (x,v)\mapsto (u,V)=(x,-1/v),
\end{equation}
we see that
$V=-\frac{U_N(x,-s)U_{N-1}(x,-s-1)}{U_{N}(x,-s-1)U_{N-1}(x,-s)}$ also
satisfies P$_{\rm III}$ with $(\theta_\infty,\theta_0)=(-s-N,-s+N-1)$.
Since it is clear that $v$ and $V$ give different functions, we obtain
two rational solutions for each $(\theta_\infty,\theta_0)=(-s-N,-s+N-1)$
which cover the case (ii).

Rational solutions for the case (i) are obtained from those for the case
(ii) by the following B\"acklund transformation,
\begin{equation}
 (\theta_\infty,\theta_0)\mapsto (\overline{\theta}_\infty,\overline{\theta}_0)
=(-\theta_\infty,\theta_0),\qquad
(x,v)\mapsto (u,V)=(\xi x,\xi v),\qquad \xi=\pm i.
\end{equation}
This implies that for each integer $N$, the functions
\begin{equation}
 v = i\, \frac{U_{N}(x/i,s-1)U_{N-1}(x/i,s)}{U_N(x/i,s)U_{N-1}(x/i,s-1)},\qquad\!\!
v=-i\, \frac{U_{N}(x/i,-s)U_{N-1}(x/i,-s-1)}{U_{N}(x/i,-s-1)U_{N-1}(x/i,-s)},\!\!
\end{equation}
give two rational solutions for $(\theta_\infty,\theta_0)=(-s+N-1,s+N)$,
which cover the case~(i).

Summarizing the discussions above, we arrive at the following conclusion:
\begin{theorem}
Let $U_N(x,s)$ ($N\in\mathbb{Z}$) be polynomials in $x$ defined by
equations~(\ref{Laguerre}), (\ref{p3tau}) and (\ref{negative}). Then, all the
 rational solutions for P$_{\rm III}$ are given as follows:

1. For any integer $N$,
\begin{equation}
 v = \frac{U_{N}(x,s-1)U_{N-1}(x,s)}{U_N(x,s)U_{N-1}(x,s-1)},\qquad
 v = -\frac{U_{N}(x,-s)U_{N-1}(x,-s-1)}{U_{N}(x,-s-1)U_{N-1}(x,-s)},
\end{equation}
give two rational solutions of P$_{\rm III}$ with
$(\theta_\infty,\theta_0)=(-s-N,-s+N-1)$.

2. For any integer $N$,
\begin{gather}
 v = i\, \frac{U_{N}(x/i,s-1)U_{N-1}(x/i,s)}
{U_N(x/i,s)U_{N-1}(x/i,s-1)},\nonumber\\
v=-i\,\frac{U_{N}(x/i,-s)U_{N-1}(x/i,-s-1)}
{U_{N}(x/i,-s-1)U_{N-1}(x/i,-s)},
\end{gather}
give two rational solutions of P$_{\rm III}$ with
$(\theta_\infty,\theta_0)=(s+N,-s+N-1)$.

3. For any integers $M$ and $N$,
\begin{gather}
  v = \frac{U_{N}(x,M-1)U_{N-1}(x,M)}{U_N(x,M)U_{N-1}(x,M-1)},\nonumber\\
 v = -\frac{U_{N}(x,-M)U_{N-1}(x,-M-1)}{U_{N}(x,-M-1)U_{N-1}(x,-M)},
\nonumber\\
v = i\,\frac{U_{M}(x/i,-N)U_{M-1}(x/i,-N-1)}{U_{M}(x/i,-N-1)U_{M-1}(x/i,-N)},
\nonumber\\
v=-i\,\frac{U_{M}(x/i,N-1)U_{M-1}(x/i,N)}{U_{M}(x/i,N)U_{M-1}(x/i,N-1)},
\end{gather}
give four rational solutions P$_{\rm III}$ with
$(\theta_\infty,\theta_0)=(-M-N,-M+N-1)$.
\end{theorem}

\begin{remark}
The case $\gamma\delta=0$ of P$_{\rm III}$ (\ref{p3}) should be
 distinguished from other cases because of the essential difference of
 type of ``space of initial conditions''~\cite{Sakai}, namely, defining
 manifold of the equation. More precisely, P$_{\rm III}$ should be
 divided into four cases; (i)~$\alpha=0$, $\gamma=0$ (or $\beta=0$,
 $\delta=0$), (ii)~Type $D_7$ ($\gamma=0$, $\alpha\delta\neq 0$ or
 $\delta=0$, $\beta\gamma\neq 0$), (iii)~Type $D_8$ ($\gamma=\delta=0$,
 $\alpha\beta\neq 0$), (iv)~Type $D_6$ (generic case)

First, in the case (i) P$_{\rm III}$ is known to be solvable by
quadratures~\cite{Gromak1,Luk1,Okamoto:p3}. Second,
type~$D_8$ does not admit any classical solutions except for two
trivial constant solutions~\cite{Ohyama}.  Also, type $D_7$ does not
admit transcendental classical solutions~\cite{Ohyama,UW:p3I}. Algebraic
 solutions for type $D_7$ have been studied by many
 authors~\cite{Gromak1,Gromak2,Luk1,Luk2,Ohyama}, but we do not discuss
 this class of solutions in this paper.
\end{remark}

\section{Rational solutions of $\boldsymbol{q}$-P$\boldsymbol{{}_{\rm III}}$}

In this section we construct a class of rational solutions of $q$-P$_{\rm III}$.
For notational simplicity, we introduce
\begin{equation}
 z = a_0q^n,\qquad y = a_0a_1q^\nu,
\end{equation}
and write equation (\ref{qP3}) as
\begin{gather}
 f_1(qz;y,c)=\frac{c^2}{f_0(z;y,c)f_1(z;y,c)}
\; \frac{1+zf_0(n;\nu,c)}{z+f_0(n;\nu,c)},\nonumber\\
 f_0(z/q;y,c)=\frac{c^2}{f_0(z;y,c)f_1(z;y,c)}
\; \frac{y/z+f_1(z;y,c)}{1+y/zf_1(z;y,c)}.\label{qP3:xy}
\end{gather}
A B\"acklund transformation is presented in \cite{KK:qp3,KNY:qp4} as
\begin{gather}
f_0(z;y,qc)=y\,f_1(z;y,c)\;
\frac{1+q/y\,f_2(n;\nu,c)+qz/y\, f_2(z;y,c)f_0(z;y,c)}{
1+z\, f_0(z;y,c)+y\,f_0(z;y,c)f_1(z;y,c)},\nonumber\\
f_1(z;y,qc)=q/z\, f_2(z;y,c)\;
\frac{1+z\, f_0(n;\nu,c)+y\,f_0(z;y,c)f_1(z;y,c)}{
1+y/z\, f_1(z;y,c)+q/z\, f_1(z;y,c)f_2(z;y,c)},\label{qP4:eq}
\end{gather}
where $f_2(z;y,c)=c^2/(f_0(z;y,c)f_1(z;y,c))$.

Now we see that equation~(\ref{qP3:xy}) admits a trivial solution,
\begin{equation}
 f_0=f_1=1,\qquad c=1.
\end{equation}
Applying the B\"acklund transformation (\ref{qP4:eq}), we observe
that
\begin{gather}
 f_0(z;y,q)=\frac{y+qz+q}{1+y+z}=q\frac{\psi_1(y/q,z)}{\psi_1(y,z)},\nonumber\\
 f_1(z;y,q)=q\frac{1+y+z}{y+z+q}=\frac{\psi_1(y,z)}{\psi_1(y/q,z/q)},
\end{gather}
where
\begin{equation}
 \psi_1(y,z)=y+z+1,
\end{equation}
is also a solution for $c=q$. Similarly, one can construct higher order
rational solutions as,
\begin{equation}
f_0\left(z;y,q^2\right) = q^2\frac{\psi_1(y,z)\psi_2(y/q,z)}{\psi_1(y/q,z)\psi_2(y,z)},\qquad
f_1\left(z;y,q^2\right) = \frac{\psi_1(y/q,z/q)\psi_2(y,z)}{\psi_1(y,z)\psi_2(y/q,z/q)},
\end{equation}
\begin{equation}
f_0\left(z;y,q^3\right) = q^2\frac{\psi_2(y,z)\psi_3(y/q,z)}{\psi_2(y/q,z)\psi_3(y,z)},\qquad
f_1\left(z;y,q^3\right) = \frac{\psi_2(y/q,z/q)\psi_3(y,z)}{\psi_2(y,z)\psi_3(y/q,z/q)},
\end{equation}
where $\psi_2(y,z)$ and $\psi_3(y,z)$ are polynomials in $y$ and $z$
given by
\begin{gather}
 \psi_2(y,z)= y^3 + \left(q+1+q^{-1}\right)y^2z +
\left(q+1+q^{-1}\right)yz^2+z^3\nonumber\\
\phantom{\psi_2(y,z)= }{}+ \left(q+1+q^{-1}\right)\left(y^2+2yz+z^2\right)
+\left(q+1+q^{-1}\right)(y+z)+1,
\\
\psi_3(y,z)=
  y^6
+ y^5z    \left(q + 1 + q^{-1}\right)\left(q + q^{-1}\right)\nonumber\\
\phantom{\psi_3(y,z)=}{}+y^4z^2  \left(q^2 + q + 1 + q^{-1} + q^{-2}\right)
\left(q + 1 + q^{-1}\right)\nonumber\\
\phantom{\psi_3(y,z)=}{}+2y^3z^3\left(q^2 + q + 1 + q^{-1} + q^{-2}\right)\left(q + q^{-1}\right)\nonumber\\
\phantom{\psi_3(y,z)=}{}+y^2z^4  \left(q^2 + q + 1 + q^{-1} + q^{-2}\right)\left(q + 1 + q^{-1}\right)\nonumber\\
\phantom{\psi_3(y,z)=}{}+ yz^5    \left(q + 1 + q^{-1}\right)\left(q + q^{-1}\right)
+ z^6\nonumber\\
\phantom{\psi_3(y,z)=}{}+ \left(q + 1 + q^{-1}\right)
 \left\{ \left(y^5(q + q^{-1}\right)+ 2y^4z\left(q^2 + q + 1 + q^{-1} + q^{-2}\right)
\right.\nonumber\\
\phantom{\psi_3(y,z)=+}{}
+y^3z^2\left(q^2 + q + 1 + q^{-1} + q^{-2}\right)\left(q^{1/2} + q^{-1/2}\right)^2\nonumber\\
\phantom{\psi_3(y,z)=+}{}
+y^2z^3\left(q^2 + q + 1 + q^{-1} + q^{-2}\right)\left(q^{1/2} + q^{-1/2}\right)^2\nonumber\\
\phantom{\psi_3(y,z)=+}{}\left.+\; 2yz^4
\left(q^2 + q + 1 + q^{-1} + q^{-2}\right)+z^5\left(q + q^{-1}\right)\right\}\nonumber\\
\phantom{\psi_3(y,z)=}{}
+\left(q^2 + q + 1 + q^{-1} + q^{-2}\right)\left(q + 1 + q^{-1}\right)\nonumber\\
\phantom{\psi_3(y,z)=+}{}\times\left\{ y^4+ y^3z \left(q^{1/2} + q^{-1/2}\right)^2
+ 2y^2z^2\left(q + 1 + q^{-1}\right)\right.\nonumber\\
\phantom{\psi_3(y,z)=+}{}\left.+ \; yz^3    \left(q^{1/2} + q^{-1/2}\right)^2+ z^4\right\}
\nonumber\\
\phantom{\psi_3(y,z)=}{}+ \left(q^2 + q + 1 + q^{-1} + q^{-2}\right)\nonumber\\
\phantom{\psi_3(y,z)=+}{}\times\left\{2y^3\left(q + q^{-1}\right)
+ y^2z\left(q + 1 + q^{-1}\right)\left(q^{1/2} + q^{-1/2}\right)^2\right.\nonumber\\
\phantom{\psi_3(y,z)=+}{}\left.+ \;yz^2\left(q + 1 + q^{-1}\right)\left(q^{1/2} + q^{-1/2}\right)^2
+ 2z^3\left(q + q^{-1}\right)\right\}\nonumber\\
\phantom{\psi_3(y,z)=}{}+ \left(q^2 + q + 1 + q^{-1} + q^{-2}\right)
\left(q + 1 + q^{-1}\right)\left(y^2+2yz+z^2\right)\nonumber\\
\phantom{\psi_3(y,z)=}{}+ \left(q + 1 + q^{-1}\right)\left(q + q^{-1}\right)(y+z)+1 ,
\end{gather}
respectively. In general, we see that $f_0\left(z;y,q^N\right)$ and $f_1\left(z;y,q^N\right)$
($N\in\mathbb{Z}_{>0}$) may be factorized as
\begin{gather}
 f_0\left(z;y,q^N\right)=q^{N} \frac{\psi_{N-1}(y,z)\psi_{N}(y/q,z)}
{\psi_{N-1}(y/q,z)\psi_{N}(y,z)},\nonumber\\
 f_1\left(z;y,q^N\right)=
\frac{\psi_{N-1}(y/q,z/q)\psi_{N}(y,z)}
{\psi_{N-1}(y,z)\psi_{N}(y/q,z/q)},
\end{gather}
respectively, where $\psi_N(y,z)$ is a polynomial in $y$ and $z$ with the
following nice properties: (1) $\psi_N(y,z)$ is a polynomial of degree
$N(N+1)/2$; (2) $\psi_N(y,z)$ is symmetric with respect to $y$ and $z$;
(3) $\psi_N(y,z)$ is symmetric with respect to $q$ and $q^{-1}$.

Now we present a determinant formula of Jacobi--Trudi type for the above
polynomials.
\begin{theorem}
\label{theorem:main}
Let  $p_k(y,z)$ $(k\in \mathbb{Z})$ be polynomials in $y$ and
$z$ defined by
\begin{equation}
\label{p:def}
 \sum_{n=0}^\infty p_n(y,z)t^n
 =\frac{(-(1-q)t;q)_\infty}
{((1-q)yt;q)_\infty ((1-q)zt;q)_\infty} ,\qquad
p_k(y,z)=0,\quad \mathrm{for}\ k<0,
\end{equation}
where $(a;q)_\infty$ is given by
\begin{equation}
 (a;q)_{\infty}=\prod_{i=0}^\infty \left(1-aq^i\right).
\end{equation}
For each $N\in\mathbb{Z}$, we define a polynomial $\phi_N(y,z)$ by
\begin{equation}
\label{phi}
 \phi_N(y,z)=
\left\{
\begin{array}{ll}
\left|
\begin{array}{cccc}
 p_N(y,z)&p_{N+1}(y,z) &\cdots &p_{2N-1}(y,z) \\
 p_{N-2}(y,z)&p_{N-1}(y,z) &\cdots &p_{2N-3}(y,z) \\
 \vdots &\cdots & \ddots & \vdots\\
 p_{-N+2}(y,z) & p_{-N+3}(y,z) & \cdots & p_1(y,z)
\end{array}
\right|,
&N>0, \vspace{1mm}\\
1, & N=0,\vspace{1mm}\\
(-1)^{N(N+1)/2} \phi_{-N-1}, & N<0.
\end{array}
\right.
\end{equation}
Then,
\begin{gather}
 f_0\left(z;y,q^{N}\right)=
q^{N} \frac{\phi_{N-1}(y,z)\phi_{N}(y/q,z)}{\phi_{N-1}(y/q,z)\phi_{N}(y,z)}, \nonumber\\
f_1\left(z;y,q^{N}\right)=
\frac{\phi_{N-1}(y/q,z/q)\phi_{N}(y,z)}{\phi_{N-1}(y,z)\phi_{N}(y/q,z/q)},\label{f and phi}
\end{gather}
satisfy $q$-P$_{\rm III}$ (\ref{qP3:xy}) with $c=q^{N}$.
\end{theorem}

\begin{remark}
1. Two polynomials $\psi_N(y,z)$ and $\phi_N(y,z)$ are related as
\begin{equation}
 \psi_N(y,z)=c_N \,\phi_N(y,z),\qquad c_N=q^{-\frac{(N-1)N(N+1)}{6}}
\prod_{k=1}^{N}[2k-1]!!,
\end{equation}
where
\begin{equation}
 [k]=\frac{1-q^k}{1-q},\qquad [2k-1]!!=[2k-1][2k-3]\cdots[3][1].
\end{equation}
In terms of the ``hook-length'' $h_{i,j}$ for the partition $\lambda=
(\lambda_1,\lambda_2,\ldots\!)=(N,N-1,\ldots,1)$~\cite{Mac},
the constant $c_N$ is also expressed as
\begin{equation}
 c_N = q^{-\sum_i (i-1)\lambda_i} \prod_{(i,j)\in \lambda} [h_{i,j}] .
\end{equation}

2. $\phi_N(y,z)$ is quite similar to the $q$-Schur function
    associated with two-core partition $\lambda=(N,N-1,\ldots,1)$ which
    appears in the rational solutions for $q$-KP
    hierarchy~\cite{KNY:qKP}. However, they are different due to the
       factor $(-(1-q)t;q)_\infty$ in the numerator of the generating
       function in equation~(\ref{p:def}). The first few $p_k(y,z)$ are given
       by,
\begin{gather}
 p_0(y,z)=1,\\
 p_1(y,z)=y+z+1,\\
 p_2(y,z)=\frac{y^2}{1+q}+yz+\frac{z^2}{1+q}+y+z+\frac{q}{1+q},\\
 p_3(y,z)=\frac{y^3}{(1+q+q^2)(1+q)}
+ \frac{y^2z}{1+q}+ \frac{yz^2}{1+q}
+\frac{z^3}{\left(1+q+q^2\right)(1+q)} \nonumber\\
\phantom{p_3(y,z)=}{} + \frac{y^2}{1+q}+yz+ \frac{z^2}{1+q}
 + \frac{yq}{1+q}+\frac{zq}{1+q} + \frac{q^3}{\left(1+q+q^2\right)(1+q)},\\
p_4(y,z)=
\frac{y^4}{\left(1+q+q^2+q^3\right)\left(1+q+q^2\right)(1+q)}
+ \frac{y^3z}{\left(1+q+q^2\right)(1+q)}
+\frac{y^2z^2}{(1+q)^2}\nonumber\\
\phantom{p_4(y,z)=}{}+ \frac{yz^3}{\left(1+q+q^2\right)(1+q)}
+\frac{z^4}{\left(1+q+q^2+q^3\right)\left(1+q+q^2\right)(1+q)}\nonumber\\
\phantom{p_4(y,z)=}{}+\frac{y^3}{\left(1+q+q^2\right)(1+q)}
+ \frac{y^2z}{1+q}+\frac{yz^2}{1+q}
+\frac{z^3}{\left(1+q+q^2\right)(1+q)}\nonumber\\
 \phantom{p_4(y,z)=}{}+ \frac{y^2q}{(1+q)^2}
+ \frac{yzq}{1+q} + \frac{z^2q}{(1+q)^2}\nonumber\\
 \phantom{p_4(y,z)=}{}+ \frac{yq^3}{\left(1+q+q^2\right)(1+q)}
+\frac{zq^3}{\left(1+q+q^2\right)(1+q)}\nonumber\\
\phantom{p_4(y,z)=}{}+\frac{q^6}{\left(1+q+q^2+q^3\right)\left(1+q+q^2\right)(1+q)}.
\end{gather}
Moreover, from equation~(\ref{p:def}), $p_k(y,z)$ satisfy the following
contiguity relations,
\begin{gather}
\left(1+q+\cdots+q^{k-1}\right)p_k(y,z)\nonumber\\
\qquad{}=\left(y+z+q^{k-1}\right)p_{k-1}(y,z)-(1-q)yzp_{k-2}(y,z),\label{cont1}\\
\label{cont2:z}
 p_k(y,qz)-p_k(y,z)=(q-1)zp_{k-1}(y,z),\\
\label{cont2:y}
p_k(qy,z)-p_k(y,z)=(q-1)yp_{k-1}(y,z).
\end{gather}
\end{remark}

Theorem \ref{theorem:main} is a direct consequence of the following
``multiplicative formula''.
\begin{proposition}
\label{proposition:multi}
The functions $f_0\left(z;y,q^N\right)$ and $f_1\left(z;y,q^N\right)$
defined by equation~(\ref{f and phi})
sa\-tis\-fy the following equations:
\begin{gather}
\label{mul1}
1+z\,f_0\left(z;y,q^{N}\right)=\left(1+z\right)
\frac{\phi_{N-1}(y/q,z/q)\phi_{N}(y,qz)}{\phi_{N-1}(y/q,z)\phi_{N}(y,z)},\\
\label{mul2}
1+\frac{1}{z}\,f_0\left(z;y,q^{N}\right)=q^{N}\left(1+\frac{1}{z}\right)
\frac{\phi_{N-1}(y,qz)\phi_{N}(y/q,z/q)}{\phi_{N-1}(y/q,z)\phi_{N}(y,z)},\\
\label{mul3}
1+\frac{y}{z}\, f_1\left(z;y,q^{N}\right)=
\left(1+\frac{y}{z}\right)\frac{\phi_{N-1}(y/q,z)\phi_{N}(y,z/q)}
{\phi_{N-1}(y,z)\phi_{N}(y/q,z/q)},\\
\label{mul4}
1+\frac{z}{y}\,f_1\left(z;y,q^{N}\right)=
\left(1+\frac{z}{y}\right)\frac{\phi_{N-1}(y,z/q)\phi_{N}(y/q,z)}
{\phi_{N-1}(y,z)\phi_{N}(y/q,z/q)}.
\end{gather}
\end{proposition}

In fact, it is easy to check that $q$-P$_{\rm III}$ (\ref{qP3:xy})
follows from Proposition~\ref{proposition:multi} and equation~(\ref{f and phi}).
Moreover, Proposition~\ref{proposition:multi} is derived from the bilinear difference
equations satisfied by~$\phi_N(x,y)$.
\begin{proposition}
\label{proposition:bl}
The function $\phi_N(y,z)$ defined by equations~(\ref{p:def})
and (\ref{phi}) satisfies the following bilinear difference equations:
\begin{gather}
\phi_{N+1}(y,z)\phi_N(y/q,z)+q^{N+1}z\phi_{N+1}(y/q,z)\phi_{N}(y,z)\nonumber\\
\qquad{}=(1+z)\phi_{N+1}(y,qz)\phi_{N}(y/q,z/q),\label{bl1}\\
 q^{-N-1}z\phi_{N+1}(y,z)\phi_N(y/q,z)+\phi_{N+1}(y/q,z)\phi_{N}(y,z)\nonumber\\
\qquad {}=(1+z)\phi_{N+1}(y/q,z/q)\phi_{N}(y,qz),\label{bl2}\\
 z\phi_{N+1}(y/q,z/q)\phi_N(y,z)+y\phi_{N+1}(y,z)\phi_{N}(y/q,z/q)\nonumber\\
\qquad {}=\left(y+z\right)\phi_{N+1}(y,z/q)\phi_{N}(y/q,z),\label{bl3}\\
 y\phi_{N+1}(y/q,z/q)\phi_N(y,z)+z\phi_{N+1}(y,z)\phi_{N}(y/q,z/q)\nonumber\\
\qquad {}=\left(y+z\right)\phi_{N+1}(y/q,z)\phi_{N}(y,z/q).\label{bl4}
\end{gather}
\end{proposition}

Multiplicative formulas (\ref{mul1})--(\ref{mul4}) are derived from
equations~(\ref{bl1})--(\ref{bl4}), respectively, by shifting $N$ to $N-1$, and
dividing the both sides by the first term of each equation.
Therefore, we have to prove Proposition~\ref{proposition:bl} in order
to establish Theorem~\ref{theorem:main}, which will be given in the next
section.

\section{Proof of Proposition \ref{proposition:bl}}\label{section:proof}

In the previous section, we have presented Theorem~\ref{theorem:main} and
shown that it is derived from Proposition~\ref{proposition:bl}. In this
section, we will prove Proposition~\ref{proposition:bl} by reducing the
bilinear equations (\ref{bl1})--(\ref{bl4}) to the Pl\"ucker relations,
namely, quadratic identities among determinants whose columns are
properly shifted. However, since equations~(\ref{bl1})--(\ref{bl4}) themselves
are not directly reduced to the Pl\"ucker relations, we will prove the
following set of bilinear difference equations instead.
\begin{proposition}
 \label{proposition:bl2}
The function $\phi_N(y,z)$ defined by equations~(\ref{p:def})
and (\ref{phi}) satisfies the following bilinear difference equations:
\begin{gather}
y\phi_{N+1}(y/q,z)\phi_N(y,z/q)-z\phi_{N+1}(y,z/q)\phi_N(y/q,z)\nonumber\\
\qquad{}=(y-z)\phi_{N+1}(y/q,z/q)\phi_N(y,z),\label{bl341}\\
y\phi_{N+1}(y,z/q)\phi_N(y/q,z)-z\phi_{N+1}(y/q,z)\phi_N(y,z/q)\nonumber\\
\qquad {}=(y-z)\phi_{N+1}(y,z)\phi_{N}(y/q,z/q),\label{bl342}\\
 q^{-N-1}z\phi_{N+1}(y,z)\phi_N(y/q,z)+\phi_{N+1}(y/q,z)\phi_{N}(y,z)\nonumber\\
\qquad{}=(1+z)\phi_{N+1}(y/q,z/q)\phi_{N}(y,qz),\label{bl2'}\\
\phi_{N+1}(y/q,z)\phi_{N-1}(y,z)- \phi_{N+1}(y,z)\phi_{N-1}(y/q,z)\nonumber\\
\qquad{}=(1-q)y/q \phi_N(y/q,z)\phi_N(y,z),\label{bl5}\\
\phi_{N+1}(y,z/q)\phi_{N-1}(y,z)- \phi_{N+1}(y,z)\phi_{N-1}(y,z/q)\nonumber\\
\qquad{}=(1-q)z/q\phi_N(y,z/q)\phi_N(y,z),\label{bl6}
\end{gather}
\begin{gather}
q^{2N}y\phi_{N}(y/q,z)\phi_N(y,z)+q^N(1+z)\phi_N(y,qz)\phi_N(y/q,z/q)\nonumber\\
\qquad{}=\frac{1-q^{2N+1}}{1-q}\, \phi_{N+1}(y,z)\phi_{N-1}(y/q,z)\label{bl7} .
\end{gather}
\end{proposition}

In fact, the bilinear difference equations (\ref{bl1})--(\ref{bl4}) in
Proposition~\ref{proposition:bl} are derived from equations~(\ref{bl341})--(\ref{bl7}) as follows.
Equation~(\ref{bl3}) is derived by adding equation~(\ref{bl341}) multiplied by $z$
to equation~(\ref{bl342}) multiplied by~$y$. Similarly, we get equation~(\ref{bl4})
by adding equation~(\ref{bl341}) multiplied by $y$ to equation~(\ref{bl342})
multiplied by $z$. Equation~(\ref{bl2'}) is the same as equation~(\ref{bl2}).
Finally, equation~(\ref{bl1}) is derived from equations~(\ref{bl2'})--(\ref{bl7})
as follows,
\begin{gather*}
 \phi_{N+1}(y,qz)\phi_N(y/q,z/q)\\
\quad {}=\frac{\phi_{N+1}(y,z)\phi_{N-1}(y,qz)-(1-q)z\phi_N(y,z)\phi_N(y,qz)}
{\phi_{N-1}(y,z)}\times\phi_N(y/q,z/q)\\
\quad = \frac{\phi_{N+1}(y,z)}{\phi_{N-1}(y,z)}\times
\left[\frac{q^{-N}z}{1+z}\phi_N(y,z)\phi_{N-1}(y/q,z)
+\frac{1}{1+z}\phi_N(y/q,z)\phi_{N-1}(y,z)\right]\\
\qquad {}-(1-q)z\frac{\phi_N(y,z)\phi_N(y,qz)\phi_N(y/q,z/q)}{\phi_{N-1}(y,z)}\\
\quad = \frac{\phi_{N}(y,z)}{\phi_{N-1}(y,z)}\times\frac{q^{-N}z}{1+z}\times
\left[\phi_{N+1}(y,z)\phi_{N-1}(y/q,z)\right.\\
\qquad \left.-\,(1-q)q^N(1+z)\phi_N(y,qz)\phi_N(y/q,z/q)\right]
+\frac{1}{1+z}\phi_{N+1}(y,z)\phi_N(y/q,z)\\
\quad {}= \frac{\phi_{N}(y,z)}{\phi_{N-1}(y,z)}\times\frac{q^{-N}z}{1+z}\times
\left[(1-q)q^{2N}y\phi_N(y/q,z)\phi_N(y,z)\right.\\
\qquad \left.+\, q^{2N+1}\phi_{N+1}(y,z)\phi_{N-1}(y/q,z)\right]
+\frac{1}{1+z}\phi_{N+1}(y,z)\phi_N(y/q,z)\\
\quad {}=\frac{\phi_{N}(y,z)}{\phi_{N-1}(y,z)}\times\frac{q^{N+1}z}{1+z}\times
\phi_{N+1}(y/q,z)\phi_{N-1}(y,z)+\frac{1}{1+z}\phi_{N+1}(y,z)\phi_N(y/q,z)\\
\quad =\frac{1}{1+z}\left[q^{N+1}z\phi_{N+1}(y/q,z)\phi_{N}(y,z)+
\phi_{N+1}(y,z)\phi_N(y/q,z)\right],
\end{gather*}
where we have used equation~(\ref{bl6}) with $q\to qz$ in the first equality,
equation~(\ref{bl2'}) with $N\to N-1$ in the second equality,
equation~(\ref{bl7}) in the fourth equality and equation~(\ref{bl5}) in fifth
equality, respectively.
\begin{remark}
 The set of equations equations~(\ref{bl341})--(\ref{bl6}) is invariant with
 respect to the transformation,
\begin{equation}
\label{parity}
N\mapsto M=-N,
\end{equation}
from the relation $\phi_N=(-1)^{N(N+1)/2} \phi_{-N-1}$.
Equation~(\ref{bl7}) itself is not invariant with respect to this
transformation, but if we consider the bilinear equation,
\begin{gather}
 q^{-2(N+1)}y \phi_N(y/q,z)\phi_N(y,z)
 + q^{-(N+1)}(1+z)\phi_N(y,qz)\phi_N(y/q,z/q)\nonumber\\
\qquad {}=-\frac{1-q^{-(2N+1)}}{1-q}\phi_{N-1}(y,z)\phi_{N+1}(y/q,z) ,\label{bl8}
\end{gather}
then the set of equations (\ref{bl7}) and (\ref{bl8}) is invariant.
Equation~(\ref{bl8}) is derived by combining equations~(\ref{bl5})--(\ref{bl7}).
From this symmetry of bilinear equations, we only have to prove
Proposition~\ref{proposition:bl2} for $N\in\mathbb{Z}_{\geq 0}$.
\end{remark}

Proposition~\ref{proposition:bl2} is proved by using a technique of
determinants which was used in the previous paper~\cite{KK:qp3}. Namely,
we first prepare ``difference formulas'' which relate $\phi_N(y,z)$ with
some determinants whose columns are properly shifted from original
$\phi_N(y,z)$. Then choosing suitable Pl\"ucker relations, we obtain
bilinear difference equations with the aid of the difference formulas.
For simpler application of this technique, we refer~\cite{dKP,RT} where
it is used to construct solutions for discrete KP and relativistic Toda
lattice equations, respectively.

In this section, we demonstrate the derivation of equation~(\ref{bl341}) as an
example. Since other bilinear equations in Proposition~\ref{proposition:bl2}
are derived by using similar technique, we give the complete proof in
the appendix.

Let us first introduce a notation,
\begin{gather}
 \phi_N(y,z)=\left|
\begin{array}{cccc}
 p_N(y,z)&p_{N+1}(y,z) &\cdots &p_{2N-1}(y,z) \\
 \vdots &\vdots & \ddots & \vdots\\
 p_{-N+4}(y,z) & p_{-N+5}(y,z)&\cdots &p_3(y,z)\\
 p_{-N+2}(y,z) &p_{-N+3}(y,z) &\cdots & p_1(y,z)
\end{array}
\right|\label{det0}\\
\phantom{\phi_N(y,z)}{}=\left|-\boldsymbol{N}+\boldsymbol{2}_{{y\atop z}},
-\boldsymbol{N}+\boldsymbol{3}_{{y\atop z}},\cdots,
\boldsymbol{0}_{{y\atop z}},\boldsymbol{1}_{{y\atop z}}\right|,
\label{det1}
\end{gather}
where the symbol $\boldsymbol{k}_{{y\atop z}}$ denotes the column vector
which ends with $p_k(y,z)$,
\begin{equation}
\label{difIv}
\boldsymbol{k}_{{y\atop z}}=
\left(\begin{array}{c}\vdots \\ p_{k+2}(y,z)\\ p_k(y,z)\end{array}\right).
\end{equation}
We note that the subscripts are used to describe the shift of $y$ and
$z$, and they will be suppressed when there is no shift.
Moreover, although the height of the column vectors are $N$ in
equation~(\ref{det1}), we use the same symbol for the determinant with
different size. So the height of~$\boldsymbol{k}$ should be read
appropriately case by case.

By using this notation, we present a difference formula.
\begin{lemma}[Difference Formula~I]\label{lemma:difI}
 \begin{gather}
 \left|-\boldsymbol{N}+\boldsymbol{2},-\boldsymbol{N}+\boldsymbol{3},\ldots,
\boldsymbol{0},\boldsymbol{1}\right|
=\phi_N(y,z),\label{difI0}\\
 \left|-\boldsymbol{N}+\boldsymbol{2}_{y/q},-\boldsymbol{N}+\boldsymbol{3},\ldots,
\boldsymbol{0},\boldsymbol{1}\right|
=\phi_N(y/q,z),\label{difI1}\\
 \left|-\boldsymbol{N}+\boldsymbol{2}_{z/q},
-\boldsymbol{N}+\boldsymbol{3},\ldots,\boldsymbol{0},\boldsymbol{1}\right|
=\phi_N(y,z/q),\label{difI2}\\
 \left|-\boldsymbol{N}+\boldsymbol{3}_{y/q},-\boldsymbol{N}+
\boldsymbol{3},\ldots,\boldsymbol{0},\boldsymbol{1}\right|
=(1-q)y/q\phi_N(y/q,z),\label{difI3}\\
 \left|-\boldsymbol{N}+\boldsymbol{3}_{z/q},
-\boldsymbol{N}+\boldsymbol{3},\ldots,\boldsymbol{0},\boldsymbol{1}\right|
=(1-q)z/q\phi_N(y,z/q),\label{difI4}\\
 \left|-\boldsymbol{N}+\boldsymbol{3}_{z/q},-\boldsymbol{N}+\boldsymbol{3}_{y/q},
-\boldsymbol{N}+\boldsymbol{4}
\ldots,\boldsymbol{0},\boldsymbol{1}\right|
=(1-q)/q(z-y)\phi_N(y/q,z/q).\label{difI5}
\end{gather}
\end{lemma}

\begin{proof}
Equation~(\ref{difI0}) is nothing but equation~(\ref{det1}).
We next shift $y\to y/q$ in equation~(\ref{det1}). Then,
subtracting $(k-1)$-st column multiplied by $(1-q)y/q$ from $k$-th
column for $k=N,N-1,\ldots,2$ and using the contiguity relation
(\ref{cont2:y}), we have
\begin{gather*}
\phi_N(y/q,z)=|-\boldsymbol{N}+\boldsymbol{2}_{y/q}, -\boldsymbol{N}+\boldsymbol{3}_{y/q},\ldots,
\boldsymbol{0}_{y/q},\boldsymbol{1}_{y/q}-(1-q)y/q\cdot \boldsymbol{0}_{y/q}| \\
\phantom{\phi_N(y/q,z)}{}=|-\boldsymbol{N}+\boldsymbol{2}_{y/q}, -\boldsymbol{N}+\boldsymbol{3}_{y/q},
\ldots,\boldsymbol{0}_{y/q},\boldsymbol{1}| \\
\phantom{\phi_N(y/q,z)}{}=\cdots\\
\phantom{\phi_N(y/q,z)}{}=|-\boldsymbol{N}+\boldsymbol{2}_{y/q}, -\boldsymbol{N}+\boldsymbol{3},\ldots,\boldsymbol{0},\boldsymbol{1}|,
\end{gather*}
which is equation~(\ref{difI1}). Moreover, in the last equality,
adding the second column
to the first column multiplied by $(1-q)y/q$, and using the contiguity relation~(\ref{cont2:y}),
we have
\begin{gather*}
(1-q)y/q \phi_N(y/q,z)=|(1-q)y/q\times(-\boldsymbol{N}+\boldsymbol{2}_{y/q})+(-\boldsymbol{N}+\boldsymbol{3}),
-\boldsymbol{N}+\boldsymbol{3},\ldots,\boldsymbol{0},\boldsymbol{1}| \\
\phantom{(1-q)y/q \phi_N(y/q,z)}{}=|-\boldsymbol{N}+\boldsymbol{3}_{y/q},-\boldsymbol{N}+\boldsymbol{3},\ldots,\boldsymbol{0},\boldsymbol{1}| ,
\end{gather*}
which is equation~(\ref{difI2}). Equations~(\ref{difI3}) and (\ref{difI4}) are
derived by the same procedure by using the contiguity relation
(\ref{cont2:z}). Finally, shifting $y\to y/q$ in equation~(\ref{difI4}),
subtracting $(k-1)$-st
column multiplied by $(1-q)y/q$ from $k$-th column for
$k=N,N-1,\ldots,3$ and using the contiguity relation (\ref{cont2:y}), we have
\begin{gather*}
(1-q)z/q\phi_N(y/q,z/q)
=|-\boldsymbol{N}+\boldsymbol{3}_{{y/q \atop z/q}}, -\boldsymbol{N}+\boldsymbol{3}_{y/q},-\boldsymbol{N}+\boldsymbol{4}_{y/q},\ldots,
\boldsymbol{0}_{y/q},\boldsymbol{1}_{y/q}| \\
\phantom{(1-q)z/q\phi_N(y/q,z/q)}{}=|-\boldsymbol{N}+\boldsymbol{3}_{{y/q \atop z/q}}, -\boldsymbol{N}+\boldsymbol{3}_{y/q},
-\boldsymbol{N}+\boldsymbol{4},\ldots,
\boldsymbol{0},\boldsymbol{1}|.
\end{gather*}
Multiplying the first column by $z-y$ and using the contiguity relation
which follows from equations~(\ref{cont2:y}) and (\ref{cont2:z}),
\begin{equation}
 zp_k(y,z/q)-yp_k(y/q,z)=(z-y)p_{k}(y/q,z/q),\label{cont2:yz}
\end{equation}
we obtain
\begin{gather*}
(1-q)z/q(z-y)\phi_N(y/q,z/q)\\
\qquad{}=|(z-y)\times(-\boldsymbol{N}+\boldsymbol{3}_{{y/q \atop z/q}}), -\boldsymbol{N}+\boldsymbol{3}_{y/q},-\boldsymbol{N}+\boldsymbol{4},
\ldots,\boldsymbol{0},\boldsymbol{1}| \\
\qquad{}=|z\times(-\boldsymbol{N}+\boldsymbol{3}_{z/q})-y\times(-\boldsymbol{N}+\boldsymbol{3}_{y/q}),
-\boldsymbol{N}+\boldsymbol{3}_{y/q},-\boldsymbol{N}+\boldsymbol{4},\ldots,\boldsymbol{0},\boldsymbol{1}| \\
\qquad{}=z|-\boldsymbol{N}+\boldsymbol{3}_{z/q},
-\boldsymbol{N}+\boldsymbol{3}_{y/q},-\boldsymbol{N}+\boldsymbol{4},\ldots,\boldsymbol{0},\boldsymbol{1}| ,
\end{gather*}
which is equation~(\ref{difI5}). This completes the proof of Lemma~\ref{lemma:difI}.
\end{proof}

Consider the Pl\"ucker relation,
\begin{gather}\label{pl1}
 0=\left|\boldsymbol{\varphi},-\boldsymbol{N}+\boldsymbol{2},-\boldsymbol{N}+\boldsymbol{3},\ldots,\boldsymbol{1}\right|
\times
\left|-\boldsymbol{N}+\boldsymbol{2}_{z/q},-\boldsymbol{N}+\boldsymbol{2}_{y/q},-\boldsymbol{N}+\boldsymbol{3},
\ldots,\boldsymbol{1}\right|\\
\phantom{0=}{}-\left|-\boldsymbol{N}+\boldsymbol{2}_{z/q},-\boldsymbol{N}+\boldsymbol{2},-\boldsymbol{N}+\boldsymbol{3},\ldots,\boldsymbol{1}\right|
\times \left|\boldsymbol{\varphi},-\boldsymbol{N}+\boldsymbol{2}_{y/q},-\boldsymbol{N}+\boldsymbol{3},
\ldots,\boldsymbol{1}\right|\nonumber\\
\phantom{0=}{}+\left|-\boldsymbol{N}+\boldsymbol{2}_{y/q},-\boldsymbol{N}+\boldsymbol{2},-\boldsymbol{N}+\boldsymbol{3},\ldots,\boldsymbol{1}\right|
\times \left|\boldsymbol{\varphi},-\boldsymbol{N}+\boldsymbol{2}_{z/q},-\boldsymbol{N}+\boldsymbol{3},
\ldots,\boldsymbol{1}\right|,\nonumber
\end{gather}
where $\varphi$ is the column vector,
\begin{equation}\label{varphi}
 \varphi = \left(\begin{array}{c}1\\ 0\\ \vdots\\ 0\end{array}\right).
\end{equation}
After expanding the determinants according to the column $\varphi$, we apply
Lemma~\ref{lemma:difI} to equation~(\ref{pl1}). Then we obtain,
\begin{gather*}
 0= \phi_N(y,z)\times (1-q)/q (z-y) \phi_{N+1}(y/q,z/q)\\
  \phantom{0=}{}- (1-q)z/q \phi_{N+1}(y/q,z)\times \phi_N(y/q,z)
+ (1-q)y/q \phi_{N+1}(y,z/q)\times \phi_N(y,z/q),
\end{gather*}
which yields equation~(\ref{bl341}). For the derivation of
equations~(\ref{bl342})--(\ref{bl7}), see appendix.

\section{Concluding remarks}
In this and previous~\cite{KK:qp3} papers, we have considered
$q$-P$_{\rm III}$ and shown that it has the following properties:
\begin{itemize}
 \item $q$-P$_{\rm III}$ is derived from the (generalization of)
       discrete-time relativistic Toda lattice.
 \item $q$-P$_{\rm III}$ and $q$-P$_{\rm IV}$ are realized as the
       dynamical systems on the root lattice of type $A^{(1)}_1\times
       A^{(1)}_2$. The former equation describes the B\"acklund
       (Schlesinger) transformation of the latter, and vice versa.
 \item $q$-P$_{\rm III}$ admits symmetry of affine Weyl group
       of type $A^{(1)}_1\times A^{(1)}_1$ as the group of
       B\"acklund transformations.
 \item $q$-P$_{\rm III}$ admits two classes of Riccati type
       solutions. One class consists of such solutions that are
       expressed by Jackson's $q$-modified Bessel functions. These
       solutions are also the solutions for $q$-P$_{\rm IV}$. Another
       class consists of $q$-P$_{\rm III}$ specific solutions. Those
       classes of solutions admit determinant formula of Hankel or
       Toeplitz type.
 \item $q$-P$_{\rm III}$ admits rational solutions which are expressed
       by ratio of some subtraction-free special polynomials. Those
       special polynomials admit determinant formula of Jacobi--Trudi
       type associated with two-core partition.
\end{itemize}

It might be an interesting problem to find connections or applications
of $q$-P$_{\rm III}$ to other fields of mathematical and physical
sciences, such as $q$-orthogonal polynomials, matrix integrations or
discrete geometry. Moreover, there are many discrete Painlev\'e
equations to be studied in Sakai's classification~\cite{Sakai}.  It
might be an important problem to consider their solutions, in
particular, solutions for the equations with larger symmetries than
$q$-P$_{\rm VI}$, which are expected to be beyond the hypergeometric
world~\cite{RGTT}.

\subsection*{Acknowledgements}
The author would like to express his sincere thanks to Prof. M~Noumi,
Prof. Y~Yamada and Dr. T~Masuda for discussions, encouragement and
advice.

\appendix
\section{Proof of bilinear equations}
In this appendix, we give the data which are necessary for proving the
bilinear equations in Proposition~\ref{proposition:bl2}. Following the
procedure mentioned in
Section~\ref{section:proof}, we first give the difference formulas. We introduce
the following notations of column vectors in order to describe them:
\begin{gather}
 [\boldsymbol{k}]_y=\left(\begin{array}{c}\vdots \\ p_5\left(y,q^kz\right)\\ p_3\left(y,q^kz\right)
\\ p_1\left(y,q^kz\right)\end{array}\right),
\qquad
 [\boldsymbol{k}']_y=\left(\begin{array}{c}\vdots\\q^5 p_5\left(y,q^kz\right) \\ q^3p_3\left(y,q^kz\right)
\\ q p_1\left(y,q^kz\right)\end{array}\right),
\label{difIIv}
\\
 [\overline{\boldsymbol{k}}]_y
=\left(\begin{array}{c}\vdots \\ p_4\left(y,q^kz\right)\\
p_2\left(y,q^kz\right)\\
p_0\left(y,q^kz\right)\end{array}\right),\qquad
 [\overline{\boldsymbol{k}}']_y
=\left(\begin{array}{c}\vdots \\ q^4\frac{1-q}{1-q^3}p_4\left(y,q^kz\right)\\
q^2\frac{1-q}{1-q^2}p_2\left(y,q^kz\right)\\
q^0\frac{1-q}{1-q^1}p_0\left(y,q^kz\right)\end{array}\right).
\label{difIIIv}
\end{gather}
Note that the subscript $y$ is used for describing the shift of $y$ and
it will be suppressed when there is no shift. Moreover, although heights of
the column vectors are not specified in equations~(\ref{difIIv}) and (\ref{difIIIv}),
they should be read appropriately case by case. For example, the
symbols
$\left|[\boldsymbol{N}-\boldsymbol{1}],
[\boldsymbol{N}-\boldsymbol{2}],\ldots,[\boldsymbol{1}],[\boldsymbol{0}]\right|$
and $ \left|[\overline{\boldsymbol{N}}],[\overline{\boldsymbol{N}-\boldsymbol{1}}],\ldots,
[\overline{\boldsymbol{1}}],[\overline{\boldsymbol{0}}]\right|$ denote the
following determinants,
\begin{gather*}
 \left|[\boldsymbol{N}-\boldsymbol{1}],[\boldsymbol{N}-\boldsymbol{2}],\ldots,[\boldsymbol{1}],[\boldsymbol{0}]\right|\\
\qquad=\left|
\begin{array}{ccccc}
 p_{2N-1}\left(y,q^{N-1}z\right)& p_{2N-1}\left(y,q^{N-2}z\right)
&\cdots &p_{2N-1}(y,qz)& p_{2N-1}(y,z) \\
 p_{2N-3}\left(y,q^{N-1}z\right)& p_{2N-3}\left(y,q^{N-2}z\right) &\cdots &p_{2N-3}(y,qz)& p_{2N-3}(y,z) \\
\vdots &\vdots &\cdots & \vdots  &\vdots\\
 p_{3}\left(y,q^{N-1}z\right)& p_{3}\left(y,q^{N-2}z\right) &\cdots &p_{3}(y,qz)& p_{3}(y,z) \\
 p_{1}\left(y,q^{N-1}z\right)& p_{1}\left(y,q^{N-2}z\right) &\cdots &p_{1}(y,qz)& p_{1}(y,z)
\end{array}\right|,\\
 \left|[\overline{\boldsymbol{N}}],[\overline{\boldsymbol{N}-\boldsymbol{1}}],\ldots,
[\overline{\boldsymbol{1}}],[\overline{\boldsymbol{0}}]\right|\notag\\
\qquad{}=\left|
\begin{array}{ccccc}
 p_{2N}\left(y,q^{N}z\right)& p_{2N}\left(y,q^{N-1}z\right) &\cdots &p_{2N}(y,qz)& p_{2N}(y,z) \\
 p_{2N-2}\left(y,q^{N}z\right)& p_{2N-2}\left(y,q^{N-1}z\right) &\cdots &p_{2N-2}(y,qz)& p_{2N-2}(y,z) \\
\vdots &\vdots &\cdots & \vdots  &\vdots\\
 p_{2}\left(y,q^{N}z\right)& p_{2}\left(y,q^{N-1}z\right) &\cdots &p_{2}(y,qz)& p_{2}(y,z) \\
 p_{0}\left(y,q^{N}z\right)& p_{0}\left(y,q^{N-1}z\right) &\cdots &p_{0}(y,qz)& p_{0}(y,z)
\end{array}\right|,
\end{gather*}
respectively.

By using these notations, we present difference formulas.
\begin{lemma}[Difference Formula II]\label{lemma:difII}
\begin{gather}
\left|[\boldsymbol{N}-\boldsymbol{1}],[\boldsymbol{N}-\boldsymbol{2}],\ldots,[\boldsymbol{1}],[\boldsymbol{0}]\right|\nonumber\\
\qquad{}=(q-1)^{\frac{N(N-1)}{2}} z^{\frac{N(N-1)}{2}} q^{\frac{(N-2)(N-1)N}{6}}
\phi_N(y,z),\label{difIIfirst}\\
\left|[\boldsymbol{N}-\boldsymbol{1}],\ldots,[\boldsymbol{2}],[\boldsymbol{1}],[\boldsymbol{0}']_{y/q}\right|\nonumber\\
\qquad {}=(-1)^{N-1}(q-1)^{\frac{N(N-1)}{2}}z^{\frac{(N-1)(N-2)}{2}}
q^{\frac{N(N^2+5)}{6}} \phi_N(y/q,z),\label{difII1}\\
\left|[\boldsymbol{N}-\boldsymbol{1}],\ldots,[\boldsymbol{2}],[\boldsymbol{1}],[\boldsymbol{1}']_{y/q}\right|\nonumber\\
\qquad {}=(-1)^{N-1}(q-1)^{\frac{N(N-1)}{2}}z^{\frac{(N-1)(N-2)}{2}}q^{\frac{N(N^2+5)}{6}}
(1+qz) \phi_N(y/q,z).\label{difII2}
\end{gather}
\end{lemma}
\begin{lemma}[Difference Formula III]
 \label{lemma:difIII}
\begin{gather}
\left|[\overline{\boldsymbol{N}}],[\overline{\boldsymbol{N}-\boldsymbol{1}}],\ldots,
[\overline{\boldsymbol{1}}],[\overline{\boldsymbol{0}}]\right|
=\left\{(q-1)z\right\}^{\frac{N(N+1)}{2}}q^{\frac{(N-1)N(N+1)}{6}}\phi_N(y,z),
\label{difIIIfirst}\\
\left|[\overline{\boldsymbol{N}}],[\overline{\boldsymbol{N}-\boldsymbol{1}}],\ldots,
[\overline{\boldsymbol{1}}],[\overline{\boldsymbol{0}}'_{y/q}]\right|\nonumber\\
\quad {}=q^{\frac{N(N+1)(N+5)}{6}}y^N\{(q-1)z\}^{\frac{N(N+1)}{2}}
\prod_{j=1}^{N+1}\frac{1-q}{1-q^{2j-1}}\phi_N(y/q,z),\label{difIII1}\\
\left|[\overline{\boldsymbol{N}}],[\overline{\boldsymbol{N}-\boldsymbol{1}}],\cdots,
[\overline{\boldsymbol{1}}],[\overline{\boldsymbol{1}}'_{y/q}]\right|\nonumber\\
\quad{}=-q^{\frac{N(N+1)(N+5)}{6}}y^{N-1}\{(q-1)z\}^{\frac{N(N+1)}{2}}(1+qz)
\times\prod_{j=1}^{N+1}\frac{1-q}{1-q^{2j-1}} \phi_N(y/q,z).\label{difIII2}
\end{gather}
\end{lemma}
We prove Lemmas \ref{lemma:difII} and \ref{lemma:difIII} in the
appendix \ref{appendix:proof}.

We next give the list of the Pl\"ucker relations and difference formulas
which are necessary for the derivation of bilinear equations
(\ref{bl341})--(\ref{bl6}).  In the following, the symbol $\varphi$
denotes the column vector defined in equation~(\ref{varphi}).

\medskip

\noindent
1. Equation~(\ref{bl341})\\
{\bf Pl\"ucker relation}
\begin{gather}
0=\left|\boldsymbol{\varphi},-\boldsymbol{N}+\boldsymbol{2},-\boldsymbol{N}+\boldsymbol{3},\ldots,\boldsymbol{1}\right|
\times\left|-\boldsymbol{N}+\boldsymbol{2}_{z/q},-\boldsymbol{N}+\boldsymbol{2}_{y/q},-\boldsymbol{N}+\boldsymbol{3},
\ldots,\boldsymbol{1}\right|\nonumber\\
{}-\left|-\boldsymbol{N}+\boldsymbol{2}_{z/q},-\boldsymbol{N}+\boldsymbol{2},-\boldsymbol{N}+\boldsymbol{3},\ldots,\boldsymbol{1}\right|
\times \left|\boldsymbol{\varphi},-\boldsymbol{N}+\boldsymbol{2}_{y/q},-\boldsymbol{N}+\boldsymbol{3},
\ldots,\boldsymbol{1}\right|\nonumber\\
{}+\left|-\boldsymbol{N}+\boldsymbol{2}_{y/q},
-\boldsymbol{N}+\boldsymbol{2},-\boldsymbol{N}+\boldsymbol{3},\ldots,\boldsymbol{1}\right|
\times \left|\boldsymbol{\varphi},-\boldsymbol{N}+\boldsymbol{2}_{z/q},
-\boldsymbol{N}+\boldsymbol{3},
\ldots,\boldsymbol{1}\right|.
\end{gather}
{\bf Difference formula}  Lemma~\ref{lemma:difI}.

\medskip

\noindent
2. Equation~(\ref{bl342})\\
{\bf Pl\"ucker relation}
\begin{gather}
0=\left|-\boldsymbol{N}+\boldsymbol{1},-\boldsymbol{N}+\boldsymbol{2},\ldots,\boldsymbol{0},\boldsymbol{1}\right|
\times \left|-\boldsymbol{N}+\boldsymbol{1}_{z/q},-\boldsymbol{N}+\boldsymbol{1}_{y/q},-\boldsymbol{N}+\boldsymbol{2},
\ldots,\boldsymbol{0}\right|\nonumber\\
{} -\left|-\boldsymbol{N}+\boldsymbol{1}_{z/q},-\boldsymbol{N}+\boldsymbol{1},-\boldsymbol{N}+\boldsymbol{2},\ldots,\boldsymbol{0}\right|
\times \left|-\boldsymbol{N}+\boldsymbol{1}_{y/q},-\boldsymbol{N}+\boldsymbol{2},
\ldots,\boldsymbol{0},\boldsymbol{1}\right|\nonumber\\
{} +\left|-\boldsymbol{N}+\boldsymbol{1}_{y/q},-\boldsymbol{N}+\boldsymbol{1},-\boldsymbol{N}+\boldsymbol{2},\ldots,\boldsymbol{0}\right|
\times \left|-\boldsymbol{N}+\boldsymbol{1}_{z/q},-\boldsymbol{N}+\boldsymbol{2},
\ldots,\boldsymbol{0},\boldsymbol{1}\right|.
\end{gather}
{\bf Difference formula}  Lemma \ref{lemma:difI}.

\medskip

\noindent
{\bf Remark.} A relation
\begin{gather}
 \phi_N(y,z)=\left|
\begin{array}{cccc}
 p_N(y,z)&p_{N+1}(y,z) &\cdots &p_{2N-1}(y,z) \\
 p_{N-2}(y,z)&p_{N-1}(y,z) &\cdots &p_{2N-3}(y,z) \\
 \vdots &\cdots & \ddots & \vdots\\
 p_{-N+2}(y,z) & \cdots & p_0(y,z) & p_1(y,z)
\end{array}
\right|\nonumber\\
\phantom{\phi_N(y,z)}{} =\left|
\begin{array}{ccccc}
 p_N(y,z)&p_{N+1}(y,z) &\cdots &p_{2N-1}(y,z)&p_{2N}(y,z) \\
 p_{N-2}(y,z)&p_{N-1}(y,z) &\cdots &p_{2N-3}(y,z)&p_{2N-2}(y,z) \\
 \vdots &\cdots & \ddots & \vdots &\vdots\\
 p_{-N+2}(y,z) & \cdots & p_0(y,z) & p_1(y,z) & p_2(y,z)\\
 p_{-N}(y,z)   & \cdots & p_{-2}(y,z) & p_{-1}(y,z) & p_0(y,z)
\end{array}
\right|\!\!\label{phibig}\\
\phantom{\phi_N(y,z)}{} =|-\boldsymbol{N}, -\boldsymbol{N}+\boldsymbol{1},\ldots, \boldsymbol{0}|.\label{det2}
\end{gather}
is used for the derivation of equation~(\ref{bl342}).

\newpage

\noindent
3. Equation~(\ref{bl2'})\\
{\bf Pl\"ucker relation}
\begin{gather}
 0=
\left|[\boldsymbol{N}],[\boldsymbol{N}-\boldsymbol{1}],\ldots,[\boldsymbol{1}],[\boldsymbol{0}']_{y/q}\right|
\times
\left|[\boldsymbol{N}-\boldsymbol{1}],\ldots,[\boldsymbol{1}],[\boldsymbol{0}],\boldsymbol{\varphi}\right|\nonumber\\
\phantom{0=}{}-
\left|[\boldsymbol{N}],[\boldsymbol{N}-\boldsymbol{1}],\ldots,[\boldsymbol{1}],\boldsymbol{\varphi}\right|
\times
\left|[\boldsymbol{N}-\boldsymbol{1}],\ldots,[\boldsymbol{1}],[\boldsymbol{0}],[\boldsymbol{0}']_{y/q}\right|
\nonumber\\
\phantom{0=}{}-
\left|[\boldsymbol{N}-\boldsymbol{1}],\ldots,[\boldsymbol{1}],[\boldsymbol{0}']_{y/q},\boldsymbol{\varphi}\right|
\times
\left|[\boldsymbol{N}],[\boldsymbol{N}-\boldsymbol{1}],\ldots,[\boldsymbol{1}],[\boldsymbol{0}]\right|.
\end{gather}
{\bf Difference formula}  Lemma \ref{lemma:difII}.

\medskip

\noindent
4. Equation~(\ref{bl5})\\
{\bf Pl\"ucker relation}
\begin{gather}
0=
\left|-\boldsymbol{N}+\boldsymbol{1}_{y/q},-\boldsymbol{N}+\boldsymbol{1},-\boldsymbol{N}+\boldsymbol{2},\ldots,\boldsymbol{0}\right|
\times
\left|-\boldsymbol{N}+\boldsymbol{2},\ldots,\boldsymbol{0},\boldsymbol{1},\boldsymbol{\varphi}\right| \nonumber\\
\phantom{0=}{}+
\left|-\boldsymbol{N}+\boldsymbol{1},-\boldsymbol{N}+\boldsymbol{2},\ldots,\boldsymbol{0},\boldsymbol{1}\right|
\times
\left|-\boldsymbol{N}+\boldsymbol{1}_{y/q},-\boldsymbol{N}+\boldsymbol{2},\ldots,\boldsymbol{0},\boldsymbol{\varphi}\right|
\nonumber \\
\phantom{0=}{}-
\left|-\boldsymbol{N}+\boldsymbol{1},-\boldsymbol{N}+\boldsymbol{2},\ldots,\boldsymbol{0},\boldsymbol{\varphi}\right|
\times
\left|-\boldsymbol{N}+\boldsymbol{1}_{y/q},-\boldsymbol{N}+\boldsymbol{2},\ldots,\boldsymbol{0},\boldsymbol{1}\right| .\label{pl?}
\end{gather}
{\bf Difference formula}  Lemma \ref{lemma:difI} with equation~(\ref{det2}).

\medskip

\noindent
5. Equation~(\ref{bl6})\\
{\bf Pl\"ucker relation}
\begin{gather}
0=
\left|-\boldsymbol{N}+\boldsymbol{1}_{z/q},-\boldsymbol{N}+\boldsymbol{1},-\boldsymbol{N}+\boldsymbol{2},\ldots,\boldsymbol{0}\right|
\times
\left|-\boldsymbol{N}+\boldsymbol{2},\ldots,\boldsymbol{0},\boldsymbol{1},\boldsymbol{\varphi}\right| \nonumber\\
\phantom{0=}{}+
\left|-\boldsymbol{N}+\boldsymbol{1},-\boldsymbol{N}+\boldsymbol{2},\ldots,\boldsymbol{0},\boldsymbol{1}\right|
\times
\left|-\boldsymbol{N}+\boldsymbol{1}_{z/q},-\boldsymbol{N}+\boldsymbol{2},\ldots,\boldsymbol{0},\boldsymbol{\varphi}\right|
\nonumber \\
\phantom{0=}{}-
\left|-\boldsymbol{N}+\boldsymbol{1},-\boldsymbol{N}+\boldsymbol{2},\ldots,\boldsymbol{0},\boldsymbol{\varphi}\right|
\times
\left|-\boldsymbol{N}+\boldsymbol{1}_{z/q},-\boldsymbol{N}+\boldsymbol{2},\ldots,\boldsymbol{0},\boldsymbol{1}\right| .
\end{gather}
{\bf Difference formula}  Lemma \ref{lemma:difI} with equation~(\ref{det2}).

\medskip

\noindent
6. Equation~(\ref{bl7})\\
{\bf Pl\"ucker relation}
\begin{gather}
 0=\left|[\overline{\boldsymbol{N}}],[\overline{\boldsymbol{N}-\boldsymbol{1}}],\ldots,
[\overline{\boldsymbol{1}}],[\overline{\boldsymbol{0}}]\right|
\times
\left|[\overline{\boldsymbol{N}-\boldsymbol{1}}],\ldots,
[\overline{\boldsymbol{1}}],
[\overline{\boldsymbol{0}}'_{y/q}],\boldsymbol{\varphi}\right|\nonumber\\
\phantom{0=}{}-
\left|[\overline{\boldsymbol{N}}],[\overline{\boldsymbol{N}-\boldsymbol{1}}],\ldots,
[\overline{\boldsymbol{1}}],[\overline{\boldsymbol{0}}'_{y/q}]\right|
\times
\left|[\overline{\boldsymbol{N}-\boldsymbol{1}}],\ldots,
[\overline{\boldsymbol{1}}],
[\overline{\boldsymbol{0}}],\boldsymbol{\varphi}\right|\nonumber\\
\phantom{0=}{}+
\left|[\overline{\boldsymbol{N}}],[\overline{\boldsymbol{N}-\boldsymbol{1}}],\ldots,
[\overline{\boldsymbol{1}}],\boldsymbol{\varphi}\right|
\times
\left|[\overline{\boldsymbol{N}-\boldsymbol{1}}],\ldots,
[\overline{\boldsymbol{1}}],
[\overline{\boldsymbol{0}}],[\overline{\boldsymbol{0}}'_{y/q}]\right|.
\end{gather}
{\bf  Difference formula}  Lemma \ref{lemma:difIII}.

\section{Proof of Lemmas \ref{lemma:difII} and \ref{lemma:difIII}}\label{appendix:proof}

In this appendix, we give the proofs of difference formulas
Lemmas~\ref{lemma:difII} and \ref{lemma:difIII}.


We first rewrite the Jacobi--Trudi type determinant expression for
$\phi_N(y,z)$ (\ref{det0}) in terms of the Casorati determinant in $z$
as follows.

\begin{lemma}
\label{lemma:Casorati}
 \begin{gather}
\phi_N(y,z)=\left\{(q-1)z\right\}^{-N(N-1)/2}q^{-(N-2)(N-1)N/6}\nonumber\\
{}\times
\left|
\begin{array}{ccccc}
 p_{2N-1}\left(y,q^{N-1}z\right)& p_{2N-1}\left(y,q^{N-2}z\right)
 &\cdots &p_{2N-1}(y,qz)& p_{2N-1}(y,z) \\
 p_{2N-3}\left(y,q^{N-1}z\right)& p_{2N-3}\left(y,q^{N-2}z\right) &\cdots &p_{2N-3}(y,qz)& p_{2N-3}(y,z) \\
\vdots &\vdots &\cdots & \vdots  &\vdots\\
 p_{1}\left(y,q^{N-1}z\right)& p_{1}\left(y,q^{N-2}z\right) &\cdots &p_{1}(y,qz)& p_{1}(y,z)
\end{array}\right|.\label{Casorati}
\end{gather}
\end{lemma}

\begin{proof}
In the right hand side of equation~(\ref{det0}), adding $(k+1)$-st column to
$k$-th column multiplied by $(q-1)z$ for $k=1,\ldots,N-1$ and using the
contiguity relation~(\ref{cont2:z}), we have
\begin{gather*}
 \phi_N(y,z)=\left\{(q-1)z\right\}^{-(N-1)}\\
\phantom{\phi_N(y,z)=}{}\times
\left|
\begin{array}{ccccc}
 p_{N+1}(y,qz)& p_{N+2}(y,qz) &\cdots &p_{2N-1}(y,qz)& p_{2N-1}(y,z) \\
 p_{N-1}(y,qz)& p_{N}(y,qz) &\cdots &p_{2N-3}(y,qz)& p_{2N-3}(y,z) \\
\vdots &\vdots &\cdots & \vdots  &\vdots\\
 p_{-N+3}(y,qz)& p_{-N+4}(y,qz) &\cdots &p_{1}(y,qz)& p_{1}(y,z)
\end{array}\right|.
\end{gather*}
Moreover, adding $(k+1)$-st column to $k$-th column multiplied by $(q-1)qz$ for
$k=1,\ldots,N-2$, and continuing this procedure, we finally obtain,
\begin{gather*}
\phi_N(y,z)=\left\{(q-1)z\right\}^{-(N-1)}\times \left\{(q-1)qz\right\}^{-(N-2)}
\times\cdots\times \left\{(q-1)q^{N-2}z\right\}^{-1}\\
{}\times
\left|
\begin{array}{ccccc}
 p_{2N-1}\left(y,q^{N-1}z\right)& p_{2N-1}\left(y,q^{N-2}z\right) &\cdots &p_{2N-1}(y,qz)& p_{2N-1}(y,z) \\
 p_{2N-3}\left(y,q^{N-1}z\right)& p_{2N-3}\left(y,q^{N-2}z\right) &\cdots &p_{2N-3}(y,qz)& p_{2N-3}(y,z) \\
\vdots &\vdots &\cdots & \vdots  &\vdots\\
 p_{1}\left(y,q^{N-1}z\right)& p_{1}\left(y,q^{N-2}z\right) &\cdots &p_{1}(y,qz)& p_{1}(y,z)
\end{array}\right|,
\end{gather*}
which is desired result.
\end{proof}

We put
\begin{gather}
 \tilde{\phi}_N(y,z)=\left|
\begin{array}{ccccc}
 p_{2N-1}\left(y,q^{N-1}z\right)& p_{2N-1}\left(y,q^{N-2}z\right) &\cdots &p_{2N-1}(y,qz)& p_{2N-1}(y,z) \\
 p_{2N-3}\left(y,q^{N-1}z\right)& p_{2N-3}\left(y,q^{N-2}z\right) &\cdots &p_{2N-3}(y,qz)& p_{2N-3}(y,z) \\
\vdots &\vdots &\cdots & \vdots  &\vdots\\
 p_{1}\left(y,q^{N-1}z\right)& p_{1}\left(y,q^{N-2}z\right) &\cdots &p_{1}(y,qz)& p_{1}(y,z)
\end{array}\right|\!\!\nonumber\\
\phantom{\tilde{\phi}_N(y,z)}{}=\left\{(q-1)z\right\}^{N(N-1)/2}q^{(N-2)(N-1)N/6}\phi_N(y,z).
\label{tildephi and phi}
\end{gather}

\noindent
\textbf{Proof of Lemma \ref{lemma:difII}.}
Equation~(\ref{Casorati}) is nothing but equation~(\ref{difIIfirst}).
In order to prove equations~(\ref{difII1}) and (\ref{difII2}),
we use the contiguity relation,
\begin{equation}
\label{rec2z}
 q^{-k}zp_k(y,z)+p_k(y/q,z)=(1+z)p_k(y/q,z/q),
\end{equation}
which follows from equations~(\ref{cont1})--(\ref{cont2:y}).
Using equation~(\ref{rec2z}) for the first column of
$\tilde{\phi}_N(y/q,z)$, we have
\begin{gather*}
 \tilde{\phi}_N(y/q,z)=
\left|
\begin{array}{cccc}
 p_{2N-1}\left(y/q,q^{N-1}z\right)& p_{2N-1}\left(y/q,q^{N-2}z\right) &\cdots & p_{2N-1}(y/q,z) \\
 p_{2N-3}\left(y/q,q^{N-1}z\right)& p_{2N-3}\left(y/q,q^{N-2}z\right) &\cdots & p_{2N-3}(y/q,z) \\
\vdots &\vdots &\cdots &\vdots\\
 p_{1}\left(y/q,q^{N-1}z\right)& p_{1}\left(y/q,q^{N-2}z\right) &\cdots & p_{1}(y/q,z)
\end{array}\right|
\end{gather*}
\begin{gather*}
=
\left|
\begin{array}{cc}
 -q^{-2N+1}\left(q^{N-1}z\right)p_{2N-1}\left(y,q^{N-1}z\right)+
\left(1+q^{N-1}z\right)p_{2N-1}\left(y/q,q^{N-2}z\right)
& \cdots \\
 -q^{-2N+3}\left(q^{N-1}z\right)p_{2N-3}\left(y,q^{N-1}z\right)
+\left(1+q^{N-1}z\right)p_{2N-3}\left(y/q,q^{N-2}z\right)
& \cdots \\
\vdots &\cdots \\
 -q^{-1}\left(q^{N-1}z\right)p_{1}\left(y,q^{N-1}z\right)+\left(1+q^{N-1}z\right)p_{1}\left(y/q,q^{N-2}z\right)
&\cdots
\end{array}\right|
\end{gather*}
\begin{gather*}
=
-q^{N-1}z\left|
\begin{array}{cccc}
 q^{-2N+1}p_{2N-1}\left(y,q^{N-1}z\right)& p_{2N-1}\left(y/q,q^{N-2}z\right) &\cdots& p_{2N-1}(y/q,z)  \\
 q^{-2N+3}p_{2N-3}\left(y,q^{N-1}z\right)& p_{2N-3}\left(y/q,q^{N-2}z\right) &\cdots& p_{2N-3}(y/q,z)  \\
\vdots &\vdots &\cdots &\vdots\\
 q^{-1}p_{1}\left(y,q^{N-1}z\right)& p_{1}\left(y/q,q^{N-2}z\right) &\cdots& p_{1}(y/q,z)
\end{array}\right|.
\end{gather*}
Continuing this procedure from the second column to the $(N-1)$-st
column, we have,
\begin{gather}
 \tilde{\phi}_N(y/q,z)=
(-q^{N-1}z)\cdots(-qz)\nonumber\\
\label{difII1-}
\times
\left|
\begin{array}{cccc}
q^{-2N+1}p_{2N-1}\left(y,q^{N-1}z\right)&\cdots& q^{-2N+1}p_{2N-1}(y,qz) & p_{2N-1}(y/q,z)  \\
q^{-2N+3}p_{2N-3}\left(y,q^{N-1}z\right)&\cdots& q^{-2N+3}p_{2N-3}(y,qz) & p_{2N-3}(y/q,z)  \\
\vdots &\vdots &\cdots &\vdots\\
q^{-1}p_{1}\left(y,q^{N-1}z\right)&\cdots& q^{-1}p_{1}(y,qz) & p_{1}(y/q,z)
\end{array}\right|\\
=
(-q^{N-1}z)\cdots(-qz)q^{-N^2}\nonumber\\
\times
\left|
\begin{array}{cccc}
p_{2N-1}\left(y,q^{N-1}z\right)&\cdots& p_{2N-1}(y,qz) & q^{2N-1}p_{2N-1}(y/q,z)  \\
p_{2N-3}\left(y,q^{N-1}z\right)&\cdots& p_{2N-3}(y,qz) & q^{2N-3}p_{2N-3}(y/q,z)  \\
\vdots &\vdots &\cdots &\vdots\\
p_{1}\left(y,q^{N-1}z\right)&\cdots& p_{1}(y,qz) & qp_{1}(y/q,z)
\end{array}\right|,\nonumber
\end{gather}
which yields equation~(\ref{difII1}) by noticing equation~(\ref{tildephi and phi}).
In equation~(\ref{difII1-}), multiplying the $N$-th column by $(1+qz)$ and
using equation~(\ref{rec2z}), we have
\begin{gather*}
 \tilde{\phi}_N(y/q,z)=
\left(-q^{N-1}z\right)\cdots(-qz)\; \frac{1}{1+qz}\\
\times
\left|
\begin{array}{cccc}
q^{-2N+1}p_{2N-1}\left(y,q^{N-1}z\right)&\cdots& q^{-2N+1}p_{2N-1}(y,qz) & (1+qz)p_{2N-1}(y/q,z)  \\
q^{-2N+3}p_{2N-3}\left(y,q^{N-1}z\right)&\cdots& q^{-2N+3}p_{2N-3}(y,qz) & (1+qz)p_{2N-3}(y/q,z)  \\
\vdots &\vdots &\cdots &\vdots\\
q^{-1}p_{1}\left(y,q^{N-1}z\right)&\cdots& q^{-1}p_{1}(y,qz) & (1+qz)p_{1}(y/q,z)
\end{array}\right|\\
=
\left(-q^{N-1}z\right)\cdots(-qz)\;\frac{1}{1+qz}\\
\times
\left|
\begin{array}{ccc}
\cdots& q^{-2N+1}p_{2N-1}(y,qz) & q^{-2N+1}(qz)p_{2N-1}(y,qz)+p_{2N-1}(y/q,qz)  \\
\cdots& q^{-2N+3}p_{2N-3}(y,qz) & q^{-2N+3}(qz)p_{2N-3}(y,qz)+p_{2N-3}(y/q,qz)  \\
\vdots &\cdots &\vdots\\
\cdots& q^{-1}p_{1}(y,qz) & q^{-1}(qz)p_{1}(y,qz) +p_1(y/q,qz)
\end{array}\right|\\
=
\left(-q^{N-1}z\right)\cdots(-qz)\; \frac{1}{1+qz}\times\left|
\begin{array}{ccc}
\cdots& q^{-2N+1}p_{2N-1}(y,qz) & p_{2N-1}(y/q,qz)  \\
\cdots& q^{-2N+3}p_{2N-3}(y,qz) & p_{2N-3}(y/q,qz)  \\
\vdots &\cdots &\vdots\\
\cdots& q^{-1}p_{1}(y,qz) & p_1(y/q,qz)
\end{array}\right|\\
=
\left(-q^{N-1}z\right)\cdots(-qz)\;\frac{1}{1+qz} \;q^{-N^2}\\
\times
\left|
\begin{array}{cccc}
p_{2N-1}\left(y,q^{N-1}z\right)&\cdots& p_{2N-1}(y,qz) & q^{2N-1}p_{2N-1}(y/q,qz)  \\
p_{2N-3}\left(y,q^{N-1}z\right)&\cdots& p_{2N-3}(y,qz) & q^{2N-3}p_{2N-3}(y/q,qz)  \\
\vdots&\vdots &\cdots &\vdots\\
p_{1}\left(y,q^{N-1}z\right)&\cdots& p_{1}(y,qz) & q p_1(y/q,qz)
\end{array}\right|,
\end{gather*}
which yields equation~(\ref{difII2}). This completes
 the proof of Lemma~\ref{lemma:difII}.~\hfill\qed

We finally prove Lemma \ref{lemma:difIII}. Starting from
equation~(\ref{phibig}), we rewrite it in terms of Casorati determinant in
$z$. Since this is done by using the same technique as that was used in
the proof of Lemma~\ref{lemma:Casorati}, we here give only the result.
\begin{lemma}\label{lemma:Casorati2}
\begin{gather}
\phi_N(y,z)=\left\{(q-1)z\right\}^{-N(N+1)/2}q^{-(N-1)N(N+1)/6}\nonumber\\
\times
\label{barphi0}
\left|
\begin{array}{ccccc}
 p_{2N}\left(y,q^{N}z\right)& p_{2N}\left(y,q^{N-1}z\right) &\cdots &p_{2N}(y,qz)& p_{2N}(y,z) \\
 p_{2N-2}\left(y,q^{N}z\right)& p_{2N-2}\left(y,q^{N-1}z\right) &\cdots &p_{2N-2}(y,qz)& p_{2N-2}(y,z) \\
\vdots &\vdots &\cdots & \vdots  &\vdots\\
 p_{2}\left(y,q^{N}z\right)& p_{2}\left(y,q^{N-1}z\right) &\cdots &p_{2}(y,qz)& p_{2}(y,z) \\
 p_{0}\left(y,q^{N}z\right)& p_{0}\left(y,q^{N-1}z\right) &\cdots &p_{0}(y,qz)& p_{0}(y,z)
\end{array}\right|.
\end{gather}
\end{lemma}
We put
\begin{gather}
\label{barphi}
\hspace{-6mm}\overline{\phi}_N(y,z)=
\left|\!
\begin{array}{ccccc}
 p_{2N}\left(y,q^{N}z\right)& p_{2N}\left(y,q^{N-1}z\right) &\cdots &p_{2N}(y,qz)& p_{2N}(y,z) \\
 p_{2N-2}\left(y,q^{N}z\right)& p_{2N-2}\left(y,q^{N-1}z\right) &\cdots &p_{2N-2}(y,qz)& p_{2N-2}(y,z) \\
\vdots &\vdots &\cdots & \vdots  &\vdots\\
 p_{2}\left(y,q^{N}z\right)& p_{2}\left(y,q^{N-1}z\right) &\cdots &p_{2}(y,qz)& p_{2}(y,z) \\
 p_{0}\left(y,q^{N}z\right)& p_{0}\left(y,q^{N-1}z\right) &\cdots &p_{0}(y,qz)& p_{0}(y,z)
\end{array}\!\right|\!\!\\
\label{barphi and phi}
\quad=\left\{(q-1)z\right\}^{N(N+1)/2}q^{(N-1)N(N+1)/6}\phi_N(y,z).
\end{gather}

\noindent
\textbf{Proof of Lemma \ref{lemma:difIII}.}
Equation (\ref{barphi0}) is nothing but equation (\ref{difIIIfirst}). In order to prove
equations~(\ref{difIII1}) and (\ref{difIII2}), we use the contiguity relation,
\begin{equation}
 q^{-k}\frac{1-q^{k+1}}{1-q}p_{k+1}(qy,z)=(1+z)p_{k}(y,z/q)+qyp_k(y,z),
\label{rec3z}
\end{equation}
which follows from equations~(\ref{cont1})--(\ref{cont2:y}).  In the right hand
side of equation~(\ref{barphi}), we add the second column multiplied by
$\left(1+q^Nz\right)$ to the first column multiplied by $qy$. Using
equation~(\ref{rec3z}), we have
\begin{gather*}
 \overline{\phi}_{N}(y,z)
=\left|
\begin{array}{ccccc}
 p_{2N}\left(y,q^{N}z\right)& p_{2N}\left(y,q^{N-1}z\right) &\cdots &p_{2N}(y,qz)& p_{2N}(y,z) \\
 p_{2N-2}\left(y,q^{N}z\right)& p_{2N-2}\left(y,q^{N-1}z\right) &\cdots &p_{2N-2}(y,qz)& p_{2N-2}(y,z) \\
\vdots &\vdots &\cdots & \vdots  &\vdots\\
 p_{2}\left(y,q^{N}z\right)& p_{2}\left(y,q^{N-1}z\right) &\cdots &p_{2}(y,qz)& p_{2}(y,z) \\
 p_{0}\left(y,q^{N}z\right)& p_{0}\left(y,q^{N-1}z\right) &\cdots &p_{0}(y,qz)& p_{0}(y,z)
\end{array}\right|\\
=\frac{1}{qy}
\left|
\begin{array}{ccc}
 qyp_{2N}\left(y,q^{N}z\right)+\left(1+q^Nz\right)p_{2N}\left(y,q^{N-1}z\right)& p_{2N}
\left(y,q^{N-1}z\right) &\cdots \\
 qyp_{2N-2}\left(y,q^{N}z\right)+\left(1+q^Nz\right)p_{2N-2}
\left(y,q^{N-1}z\right)& p_{2N-2}\left(y,q^{N-1}z\right) &\cdots \\
\vdots &\vdots &\cdots \\
 qyp_{0}\left(y,q^{N}z\right)+\left(1+q^Nz\right)p_{0}\left(y,q^{N-1}z\right)& p_{0}\left(y,q^{N-1}z\right) &\cdots
\end{array}\right|\\
=\frac{1}{qy}
\left|
\begin{array}{ccc}
 q^{-2N}\frac{1-q^{2N+1}}{1-q}p_{2N+1}\left(qy,q^{N}z\right)& p_{2N}\left(y,q^{N-1}z\right) &\cdots \\
 q^{-2N+2}\frac{1-q^{2N-1}}{1-q}p_{2N-1}\left(qy,q^{N}z\right)& p_{2N-2}\left(y,q^{N-1}z\right) &\cdots \\
\vdots &\vdots &\cdots \\
 q^{0}\frac{1-q^{1}}{1-q}p_{1}\left(qy,q^{N}z\right)& p_{0}\left(y,q^{N-1}z\right) &\cdots
\end{array}\right|.
\end{gather*}
Continuing this procedure from the second to the $N$-th columns,
we obtain
\begin{gather}
\label{difIII1-}
\overline{\phi}_{N}(y,z)=
\left(\frac{1}{qy}\right)^{N}
\times\left|
\begin{array}{ccc}
 \cdots &
 q^{-2N}\frac{1-q^{2N+1}}{1-q}\;p_{2N+1}(qy,qz) & p_{2N}(y,z)\\
\cdots &
 q^{-2N+2}\frac{1-q^{2N-1}}{1-q}\;p_{2N-1}(qy,qz) & p_{2N-2}(y,z)\\
\vdots &\vdots &\cdots \\
\cdots & q^{0}\frac{1-q^{1}}{1-q} \; p_{1}(qy,qz) & p_{0}(y,z)
\end{array}\right|\\
\qquad=\left(\frac{1}{qy}\right)^{N}\prod_{j=1}^{N+1}\frac{q^{-2j}(1-q^{2j-1})}{1-q}
\nonumber\\
\qquad \times\left|
\begin{array}{cccc}
p_{2N+1}\left(qy,q^{N}z\right)& \cdots &
p_{2N+1}(qy,qz) &  q^{2N}\frac{1-q}{1-q^{2N+1}}\;p_{2N}(y,z)\\
p_{2N-1}\left(qy,q^{N}z\right)& \cdots &
p_{2N-1}(qy,qz) &  q^{2N-2}\frac{1-q}{1-q^{2N-1}}\;p_{2N-2}(y,z)\\
\vdots &\vdots &\cdots \\
p_{1}\left(qy,q^{N}z\right)& \cdots &
p_{1}(qy,qz) &  q^{0}\frac{1-q}{1-q^{1}}\;p_{0}(y,z)
\end{array}\right|. \nonumber
\end{gather}
which yields equation~(\ref{difIII1}) by noticing equation~(\ref{barphi and phi}).
Furthermore, in equation~(\ref{difIII1-}), adding the $(N-1)$-st column to the
$N$-th column multiplied by $-(1+qz)$ and using equation~(\ref{rec3z}), we have
\begin{gather*}
 \overline{\phi}_{N}(y,z)=\left(\frac{1}{qy}\right)^{N}\frac{-1}{1+qz}\\
\qquad \times
\left|
\begin{array}{cccc}
\cdots & q^{-2N}\frac{1-q^{2N+1}}{1-q}p_{2N+1}(qy,qz) &
qyp_{2N}(y,qz)\\
\cdots & q^{-2N+2}\frac{1-q^{2N-1}}{1-q}p_{2N-1}(qy,qz)
&qyp_{2N-2}(y,qz)\\
\vdots &\vdots &\cdots \\
 \cdots & q^{0}\frac{1-q^{1}}{1-q}p_{1}(qy,qz) &
qyp_{0}(y,qz)
\end{array}\right|\\
\qquad =
\left(\frac{1}{qy}\right)^{N-1}\frac{-1}{1+qz}
\prod_{j=1}^{N+1}\frac{q^{-2(j-1)}(1-q^{2j-1})}{1-q}\\
\qquad \times\left|
\begin{array}{cccc}
p_{2N+1}\left(qy,q^{N}z\right)& \cdots &
p_{2N+1}(qy,qz) &  q^{2N}\frac{1-q}{1-q^{2N+1}}p_{2N}(y,qz)\\
p_{2N-1}\left(qy,q^{N}z\right)& \cdots &
p_{2N-1}(qy,qz) &  q^{2N-2}\frac{1-q}{1-q^{2N-1}}p_{2N-2}(y,qz)\\
\vdots &\vdots &\cdots \\
p_{1}\left(qy,q^{N}z\right)& \cdots &
p_{1}(qy,qz) &  q^{0}\frac{1-q}{1-q^{1}} \,p_{0}(y,qz)
\end{array}\right|,
\end{gather*}
which gives equation~(\ref{difIII2}). This completes the proof of Lemma~\ref{lemma:difIII}.\hfill\qed

\label{kajiwara-lastpage}

\end{document}